\documentclass[aip,10pt,twocolumn]{revtex4-1}
\usepackage{graphicx}
\usepackage{float}
\usepackage{subfig}
\usepackage{setspace}
\usepackage{color}
\usepackage{multirow}
\usepackage{amsmath}
\DeclareMathOperator{\Tr}{Tr}

\begin{document}
\title{Frequency lock closed-loop control of a separated flow using visual feedback}

\author{N. Gautier and J.-L. Aider}
\address{PMMH, 10 rue Vauquelin 75006 Paris, France}

\begin{abstract}
In this study, a simple model based closed-loop algorithm is used to control the separated flow downstream a backward-facing step. It has been shown in previous studies that the recirculation bubble can be minimized when exciting the shear layer at its natural Kelvin-Helmholtz instability frequency. In this experiment, the natural shedding frequency is identified through real-time analysis of 2D velocity fields. Actuation (pulsed jet) is then locked on this frequency. If flow characteristics stray too far from a set point, shedding frequency is updated and actuation changed. The present work demonstrates the efficacy and robustness of this approach in reducing recirculation while Reynolds number is randomly varied  between 1400 and 2800.
\end{abstract}

\maketitle
\section{Introduction}
The control of separated flows is of major academic and industrial interest. At the interface of control theory and fluid mechanics it is relevant in many engineering domains, such as aeronautics and combustion.
\\
Flow separation usually induces flow recirculations \cite{Armaly1983}, which can be detrimental (higher drag, lower lift) or beneficial (enhanced mixing). Because it creates a spatially bounded separation at the step edge \cite{Mazur2007,JLA2007}, the Backward-Facing Step (BFS) is a benchmark geometry for separated flows.
\\\\
Control of the flow downstream a BFS has been the subject of much research both numerically and experimentally. There are three distinct approaches to flow control: passive control which involves altering the geometry to yield the desired effect, open-loop control where power is supplied to the system to alter its operating conditions \cite{Chun1996} and closed-loop control \cite{King2005,King2007,Gautier2013control}. Passive control uses permanent actuations and cannot adapt to changes in the flow conditions \cite{duriez2009self}. Open-loop control can vary with user input and often involves an input-output map where desired outputs are matched to inputs \cite{Joseph2013}. Closed-loop control enhances open-loop control by using a feedback element to gauge the state of the flow in order to better follow commands and reject disturbances.
\\\\
This approach to control can be further declined: extremum-seeking controllers, where a cost variable is defined and minimized \cite{King2005,beaudoin2006drag}; black box control where the flow is excited in order to compute a model of the flow system without \textit{a priori} physical knowledge; model based control, where a model of the flow system is devised using physical knowledge and/or empirical data, which is then is used to compute appropriate actuation \cite{Pastoor2003}. A good overview of various control methods is given in \citet{King2007}.
\\\\
Regarding actuation, periodic forcing can trigger Kelvin-Helmholtz (KH) instabilities in the shear layer which prompts the creation of spanwise vortices. Forcing the shear layer close to its natural frequency has been shown to be most effective at reducing the recirculation bubble in separated flows, \cite{Chun1996}. Thus, dynamically identifying the natural shedding frequency  can be essential when considering closed loop control. Unfortunately, from an experimental point of view, measuring the  natural shear layer frequency can be challenging, especially with wall sensors.
\\
 The aim of this paper is to demonstrate the effectiveness of a simple model based closed-looped control scheme where the flow is actuated at its natural shedding frequency,  computed in real-time by optical means.

\section{Experimental Setup}
\subsection{Water tunnel}
Experiments were carried out in a hydrodynamic channel in which the flow is driven by gravity.
The flow is stabilized by divergent and convergent sections separated by honeycombs. The test section is $80$~cm long with a rectangular cross section $15$~cm wide and $10$~cm high.\\
 The quality of the main stream can be quantified in terms of flow uniformity and turbulence intensity.  The standard deviation $\sigma$ is computed for the highest free stream velocity featured in our experimental set-up. We obtain $\sigma = 0.059$~cm.s$^{-1}$ which corresponds to turbulence levels of $\frac{\sigma}{U_{\infty}}=0.0023$.
 \\
 The mean free stream velocity $U_{\infty}$ can go up to $22$~cm.s$^{-1}$ leading to a maximum Reynolds number $Re_h=\frac{U_{\infty}h}{\nu}=2800$.  A specific leading-edge profile is used to  smoothly start the boundary layer which then grows downstream along the flat plate, before reaching the edge of the step $33.5$~cm downstream. The boundary layer is laminar and follows a Blasius profile. More details can be found in \citet{cambonie2013experimental}.

\begin{figure}
\centering
\includegraphics[width=0.5\textwidth]{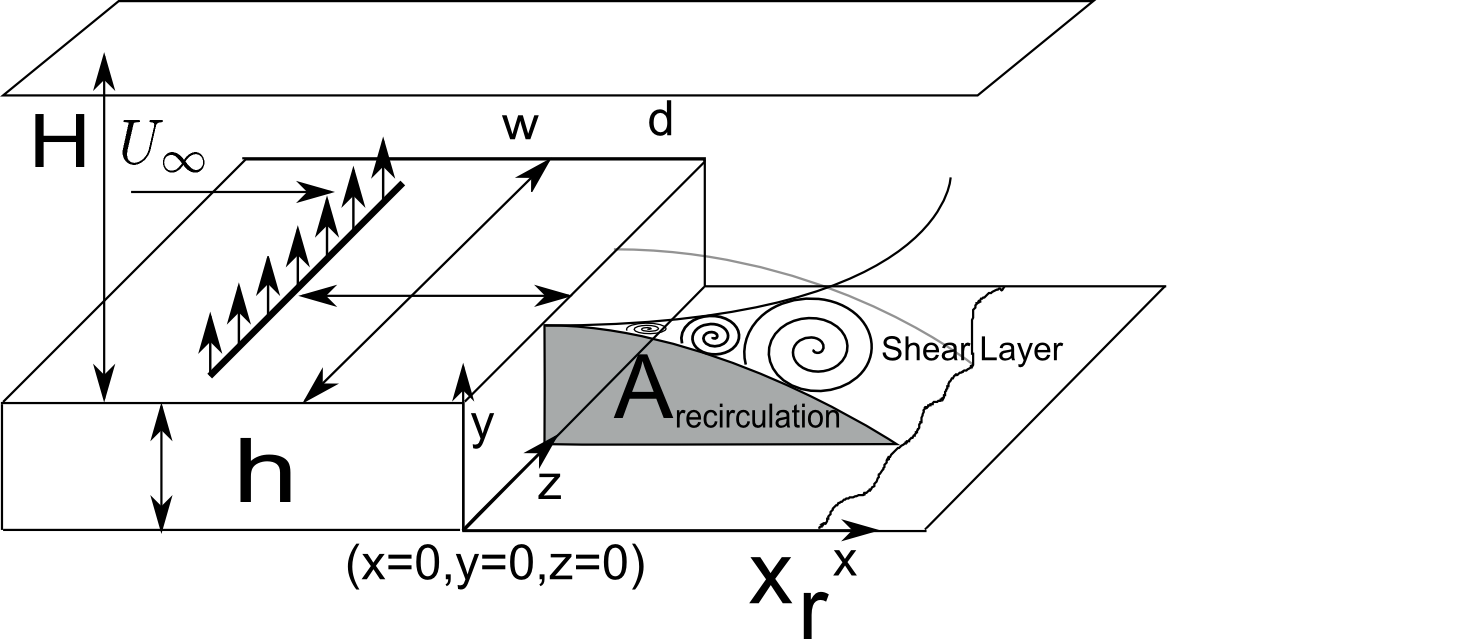}
\caption{Sketch of the BFS geometry and definition of the main parameters.}
\label{fig:dimensions}
\end{figure}

\subsection{Backward-facing step geometry}
The BFS geometry and the main geometric parameters are shown in figure~\ref{fig:dimensions}. BFS height is $h=1.5$~cm. Channel height is $H=7$~cm for a channel width $w=15$~cm. The vertical expansion ratio  is $A_y = \frac{H}{h+H} = 0.82$ and the spanwise aspect ratio is $A_z=\frac{w}{h+H}=1.76$. 

\subsection{Velocity fields computation}
The flow is seeded with 20~$\mu m$  neutrally buoyant polyamid seeding particles.  They are illuminated by a laser sheet created by a 2W continuous laser beam operating at $\lambda = 532$~nm.  Images of the symmetry plane are recorded using a Basler acA 2000-340km 8bit CMOS camera. Velocity field computations are run on a Gforce GTX 580 graphics card. \\
	
The algorithm used to compute the velocity fields is a Lukas-Kanade optical flow algorithm \cite{Champagnat2005} called FOLKI developped at ONERA. Its offline  and online accuracy have been demonstrated and detailed by \citet{Plyer2011,Gautier2013OF}. The algorithm was used off-line by \citet{Leclaire2012,Bur2012}. The GPU version (FOLKI-GPU) was improved \cite{Gautier2013OF} to allow for the computation of instantaneous velocity fields in real time at a sampling frequency $f_s$ up to 100 Hz. Thus the state of the flow can be computed in real-time up to 100 Hz and then be used as an input in closed-loop control. Another advantage of the optical flow algorithm is its ability to compute accurate velocity fields even when velocity changes by as much as a factor of 2 without adjusting acquisition or computation parameters. Fields are $17.2\times4.6$~cm$^2$ and capture the whole recirculation regardless of Reynolds numbers.

\subsection{Actuation}
Actuation is provided by a flush slot jet,  0.1~cm long and 9~cm wide. Injection is normal to the wall. The distance between the injection slot and step edge is $d=3.5$~cm~$=2.3 h$ (figure~\ref{fig:dimensions}). Water coming from a pressurized tank enters a plenum and goes through a volume of glass beads designed to homogenize the incoming flow. Jet amplitude is controlled by changing tank pressure. 
\\
The flow is modulated by a one-way voltage driven solenoid-valve. A constant amplitude square signal is sent to the valve. The duty cycle $dc$, defined as the ratio between the time during which the valve is opened $T_{on}$ and total cycle time $T_{ac}$, is kept constant at $dc = \frac{T_{on}}{T_{ac}}=20$~\% which has been shown to be an optimal value for this setup  \cite{Gautier2013upstream}. The only varying actuation feature is the frequency of the actuation $F_{ac}$.

\subsection{Flow state qualification}
The most commonly used variable to quantify the state of the flow downstream the BFS is the  length of the recirculation bubble \cite{Chun1996,King2005,King2007}. However, the recirculation area $A_{rec}$ has been shown  to be a proper state parameter, easier to compute when using 2D instantaneous velocity fields \cite{Gautier2013control,Gautier2013OF,Gautier2013upstream}. Because $A_{rec}$ is computed using the whole 2D flow field, it is able to better capture the influence of actuation on flow structures near the step edge and its influence on the whole recirculation bubble. It is computed over the 2D velocity field: 

\begin{equation}
A_{rec}(t)=\int_{A} H(-v_x) da
\label{eq:A_r}
\end{equation}

where A is the 2D velocity field area, H is the Heaviside function and $v_x$ the longitudinal velocity.

\subsection{Natural shedding frequency}
\begin{figure}
\centering
\subfloat[Contours of instantaneous $\lambda_{Ci}(x, y)$ at a given time step for $Re_h=2900$. The vertical line shows the position where the $\lambda_{Ci}(x = 5h)$ is integrated to identify the KH frequency. The red rectangle shows the position where flow velocity is computed.]{\includegraphics[width=0.45\textwidth]{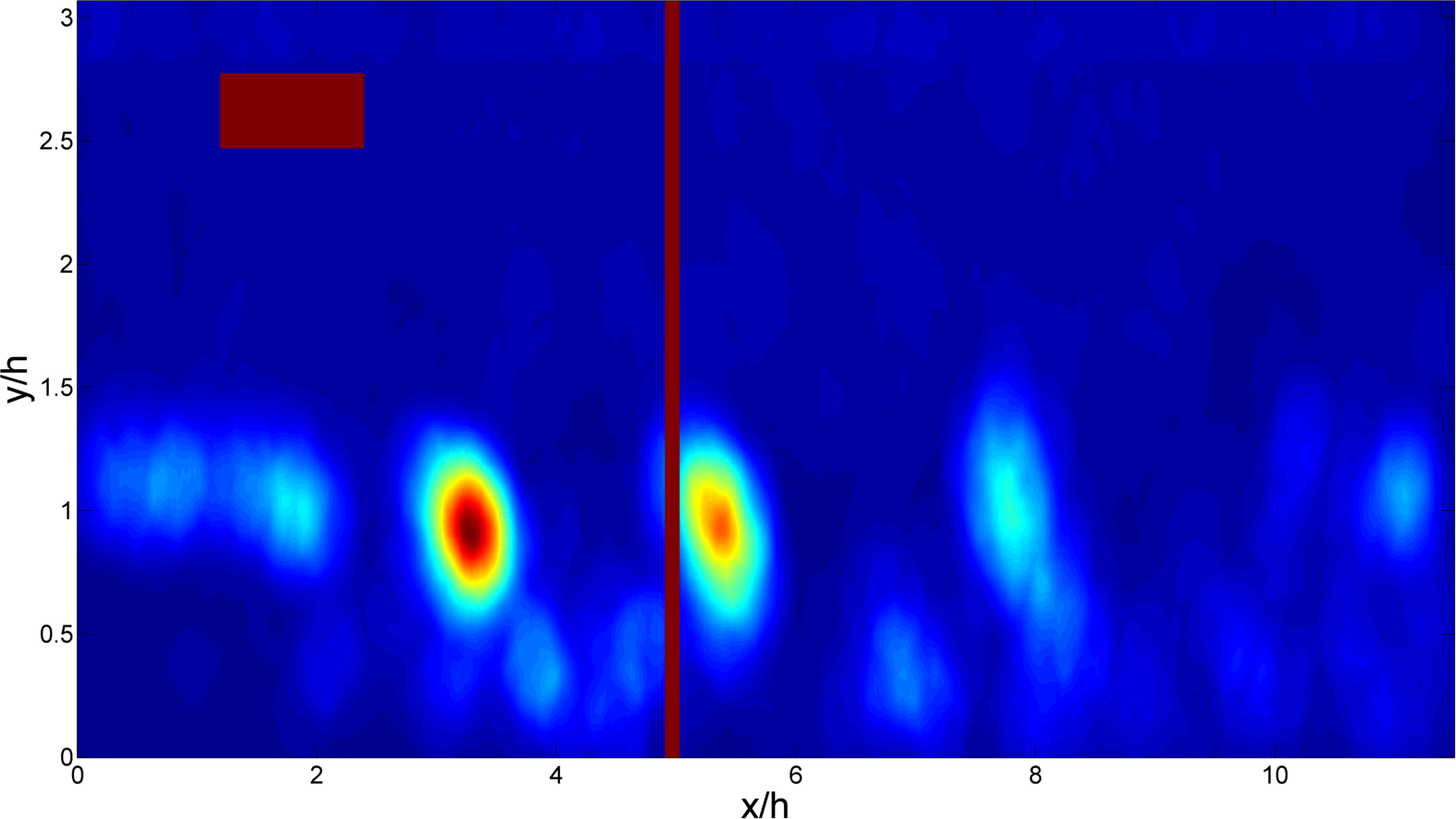}\label{fig:freq_1}}\\
\subfloat[$\lambda_{Ci}$ time series at $x=5.0h$ for $Re_h=2900$.]{\includegraphics[width=0.45\textwidth]{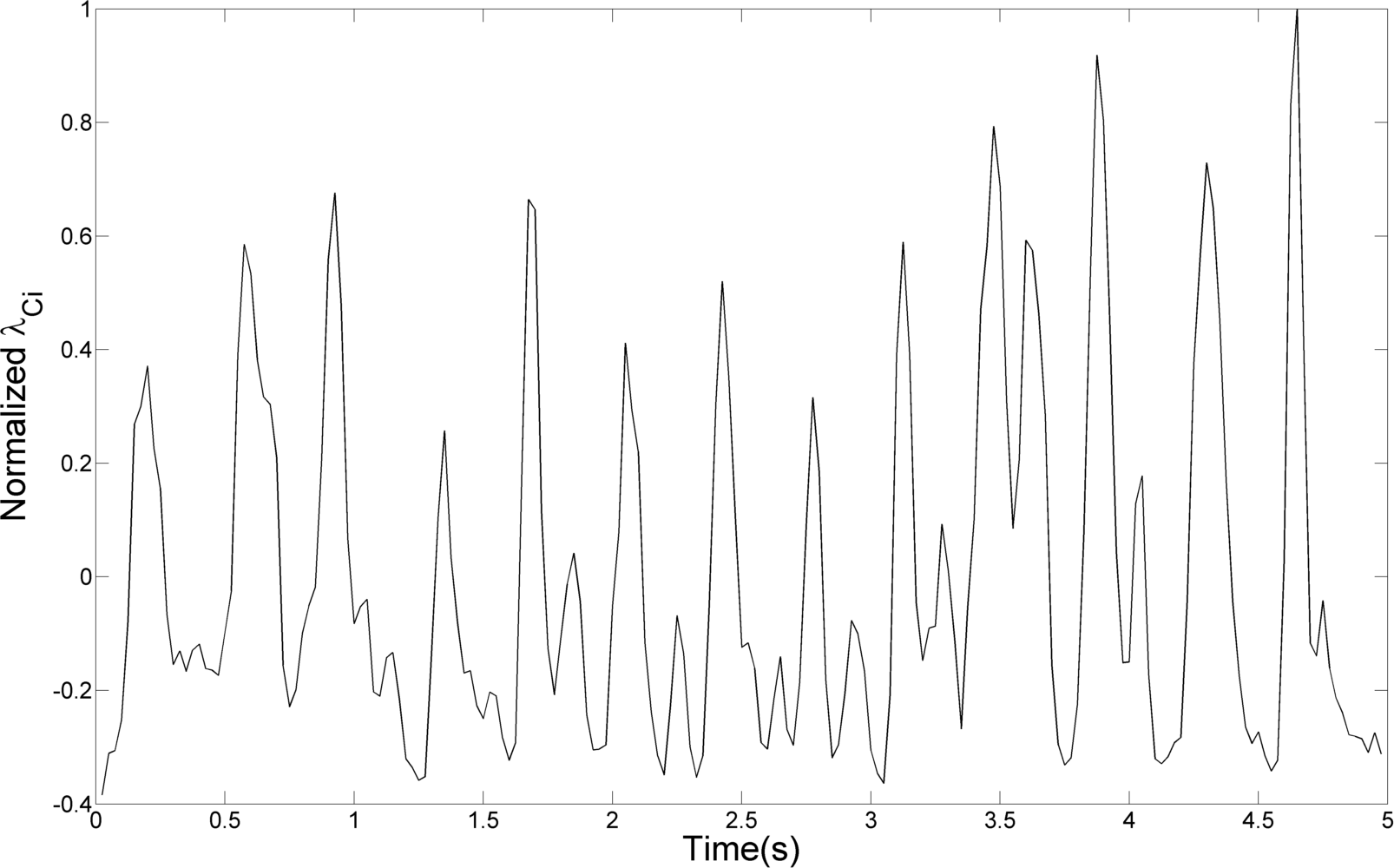}\label{fig:freq_2}}\\
\subfloat[Frequency spectrum for this time series showing a stronger peak at $f_{KH}=3.08$]{\includegraphics[width=0.45\textwidth]{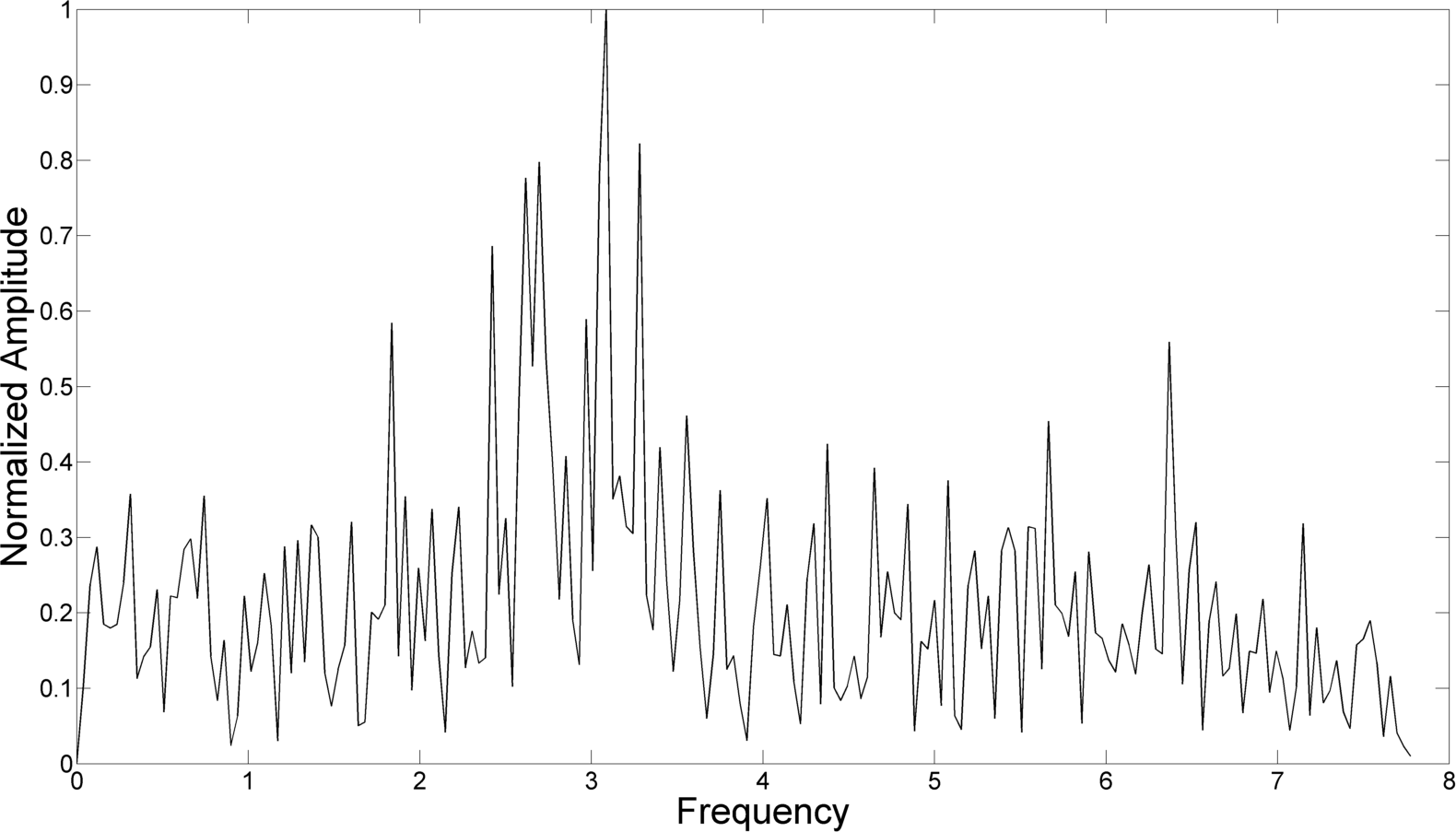}\label{fig:freq_3}}
\caption{}
\label{fig:series_freq}
\end{figure}

Kelvin-Helmholtz spanwise vortices created in the shear layer of the BFS (figure~\ref{fig:dimensions}) strongly influence the recirculation area. 
Using optical flow instantaneous velocity measurements allows the detection of KH vortices as they are shed in the shear layer.
An effective way of detecting vortices is to compute the swirling strength criterion $\lambda_{ci}(s^{-1})$ in the instantaneous two-components velocity fields.  This criterion was first introduced and subsequently improved by \citet{Chong1990,Zhou1999} who analyzed the velocity gradient tensor and proposed that the vortex core be defined as a region where $\nabla \bf{u}$ has complex conjugate eigenvalues. For 2D data $\lambda_{Ci}$ can be computed quickly and efficiently  following  equation~\ref{eq:lCi}:

\begin{equation}
\lambda_{Ci}=\frac{1}{2}\sqrt{4 \det(\nabla \bf{u})- \Tr(\nabla \bf{u})^2}
\label{eq:lCi}
\end{equation}

when such a quantity is real, else $\lambda_{Ci}=0$.  
The shedding frequency is obtained by spatially averaging $\lambda_{Ci}$ in the vertical direction at $x=5 h$. The sampling frequency is $f_s =$ 60Hz. This is equivalent to counting vortices as they pass through an imaginary vertical line (Figure~\ref{fig:freq_1}). Figure~\ref{fig:freq_2} shows a typical time series of fluctuations of $\lambda_{Ci}(t)$  for $Re_h=2900$. Every peaks correspond to a vortex moving across the $x = 5h$ line. Figure~\ref{fig:freq_3} shows the corresponding frequency spectrum obtained by Fourier transform. The natural shedding frequency $f_{KH}$ is well defined. It leads to a Strouhal number based on the step height $St_h=\frac{f_{KH} h}{U_0} =0.291$ for the KH shedding frequency. 

Because recirculation changes with $Re_h$ and control actuation it is important to choose a position where shedding frequency can be computed reliably. This is done by placing the detection zone such that initial vortex shedding position is always upstream whatever the operating conditions. In practice this is achieved by placing the detection zone sufficiently downstream the BFS, in this case at $x =5.0h$.

\section{Results}
\subsection{Control algorithm}

\begin{figure}
\centering
\includegraphics[width=0.45\textwidth]{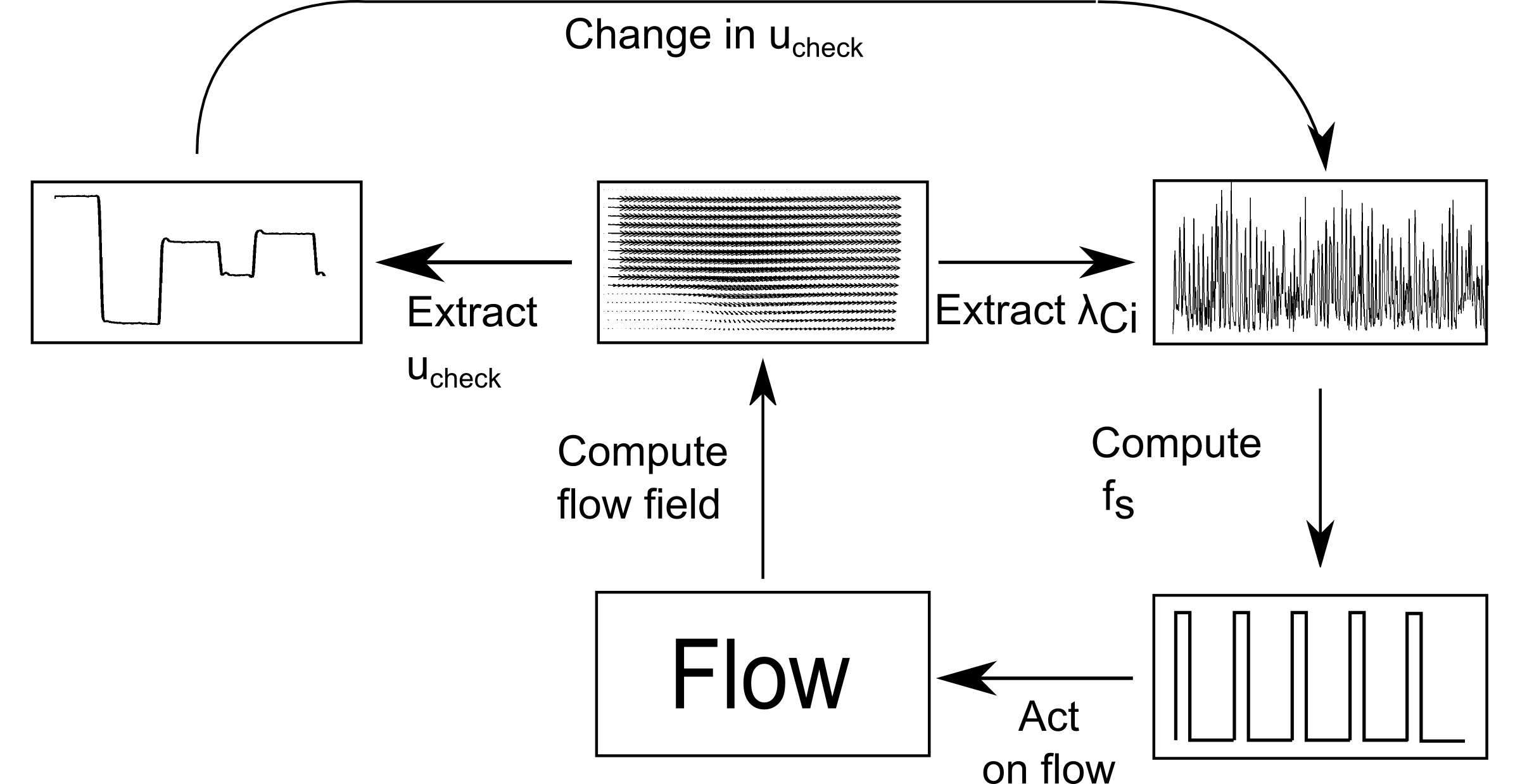}
\caption{Frequency lock algorithm}
\label{fig:control_algorithm}
\end{figure}

The closed-loop control algorithm is described in figure~\ref{fig:control_algorithm}. A new quantity, called $u_{check}(t)$, is introduced. It corresponds to the longitudinal velocity field spatially averaged in the upper corner of the velocity fields, far from the boundaries. When the external flow is stabilized, i.e. when $u_{check}(t)$ does not change for a given period $\Delta T_{steady}$, then the  shedding frequency  $f_{KH}$   is computed over a given  time $\Delta T_{computation}$. Once $f_{KH}$ has been estimated, the control begins: the jet starts pulsing at resulting $f_{ac} = f_{KH}$. At this point shedding frequency is locked to pulse frequency leading, in principle, to a minimization of the recirculation area.
\\
The actuation is kept constant as long as $\Delta u_{check}$ does not change. This is done by continuously polling the value described in equation \ref{eq:polling}. It is a measure of how much the flow has changed since the last frequency computation. If the value goes above a given threshold, frequency is re-computed, thus completing the loop. The threshold depends on the noise of the monitoring variable and should be chosen such that only significant changes prompt re-computation.\\


\begin{equation}
\Delta u_{check}(t)=\frac{u_{check}(t)-u_{check}(t-\Delta t)}{u_{check}(t-\Delta t)}
\label{eq:polling}
\end{equation}

Here $u_{check}$ is chosen to determine whether frequency can be reliably computed but many other quantities might be used as inputs.  $\Delta T_{steady}$ should be chosen such that computation starts after there are no significant changes in the monitoring variable. This can be done by taking a value greater than the characteristic time scale of the system. $\Delta T_{computation}$ must be chosen so that $f_{KH}$ can be reliably computed.  Small variations in $f_{KH}$ for the same input flow can be allowed since the flow is sensitive to actuation close to $f_{KH}$. Longer checking and computation times ensure a more reliable but less responsive system.

\subsection{Validation of the frequency-lock approach for varying Reynolds numbers}

\begin{figure}
\centering
\subfloat[Random variations of the Reynolds number $Re_h$ as a function of time. ]{\includegraphics[width=0.45\textwidth]{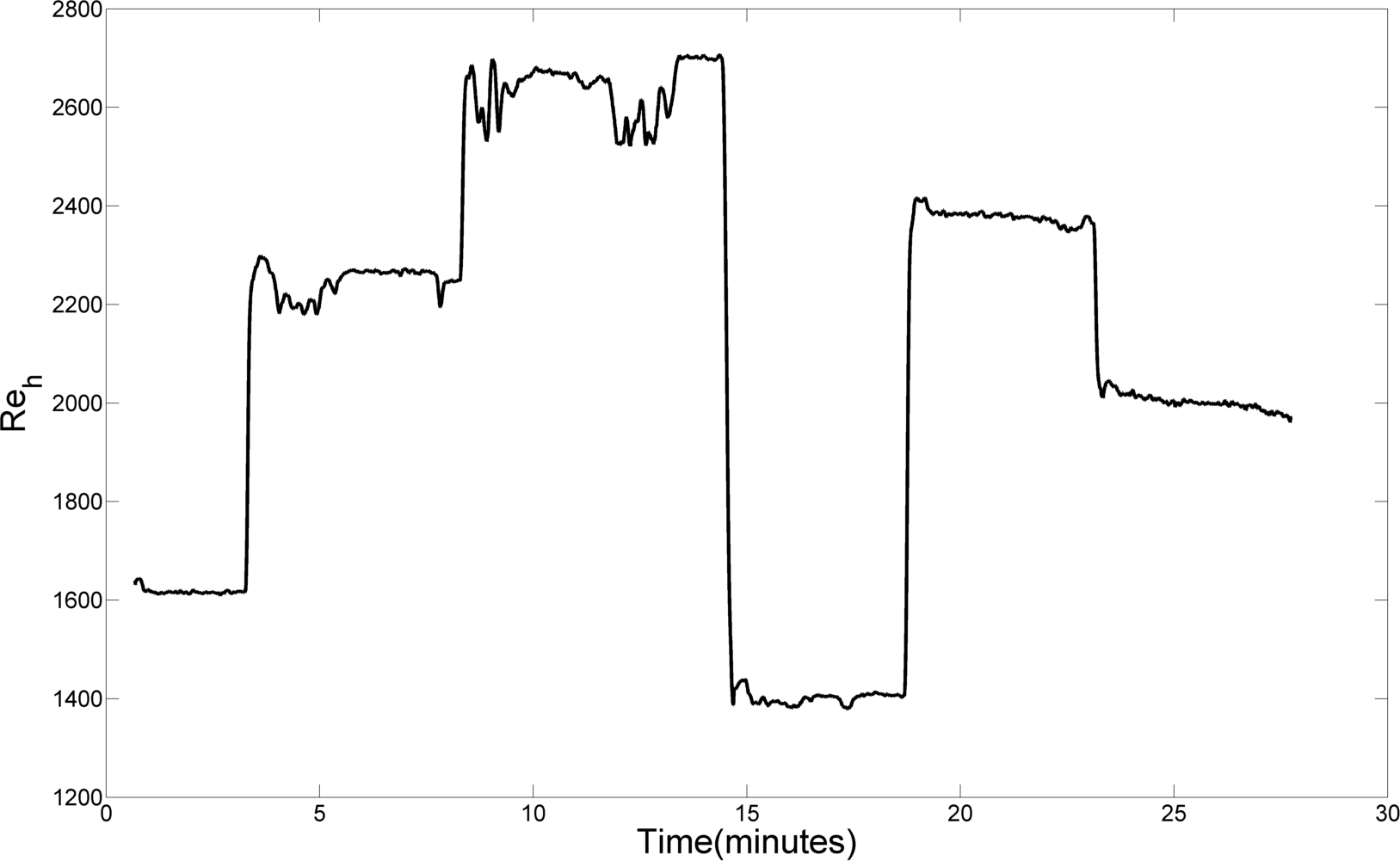}\label{fig:u_check}}\\
\subfloat[Corresponding evolution of $f_{KH}$ as a function of time, following the variations of $Re_h(t)$. ]{\includegraphics[width=0.45\textwidth]{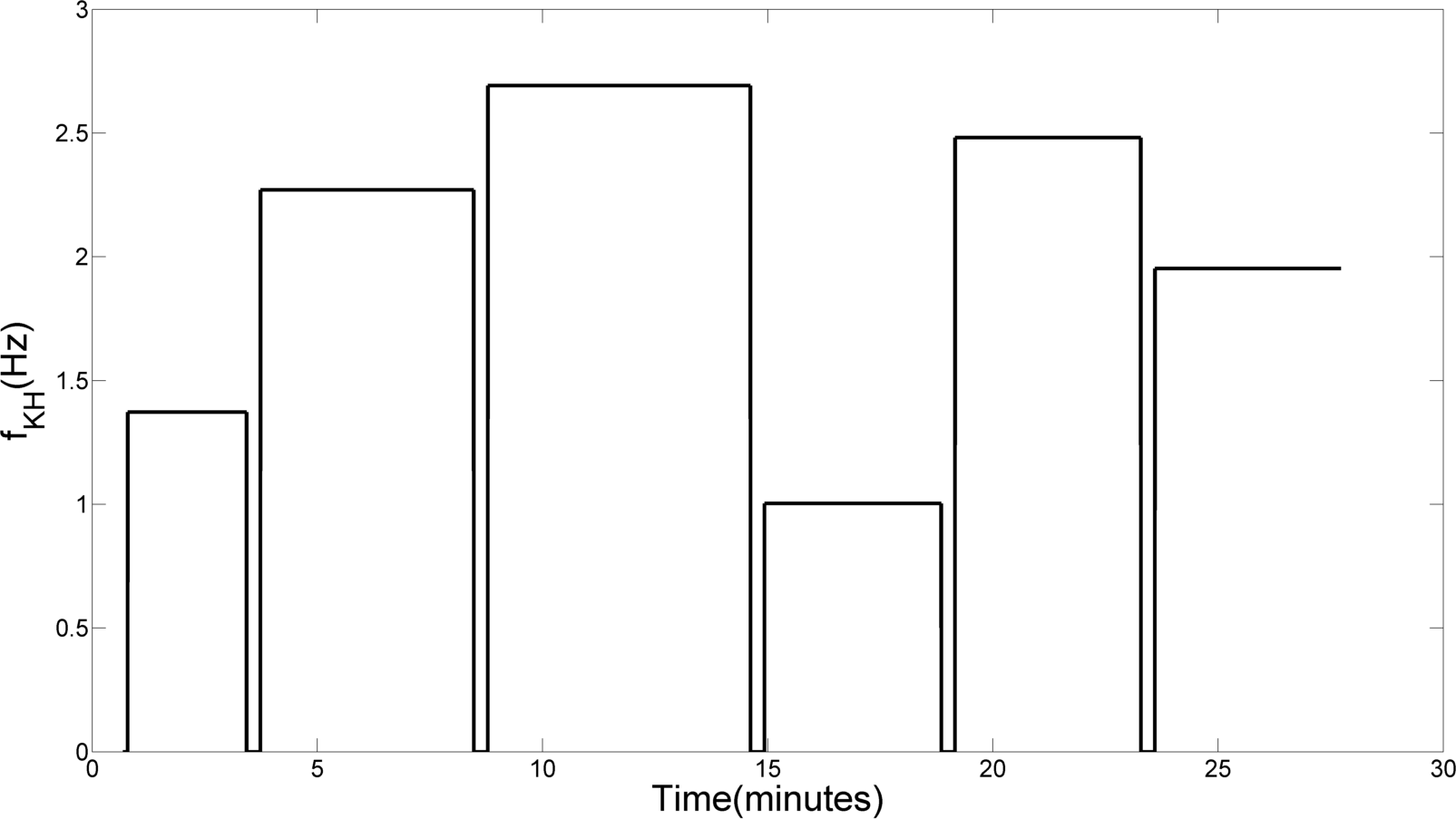}\label{fig:freq}}\\
\subfloat[Evolution of $A_{rec}/h^2$ as a function of time. Time series are normalized by the uncontrolled recirculation area corresponding to given Reynolds number. Mean values of the controlled signal are shown in red. They are computed for each period when Re is changed.]{\includegraphics[width=0.45\textwidth]{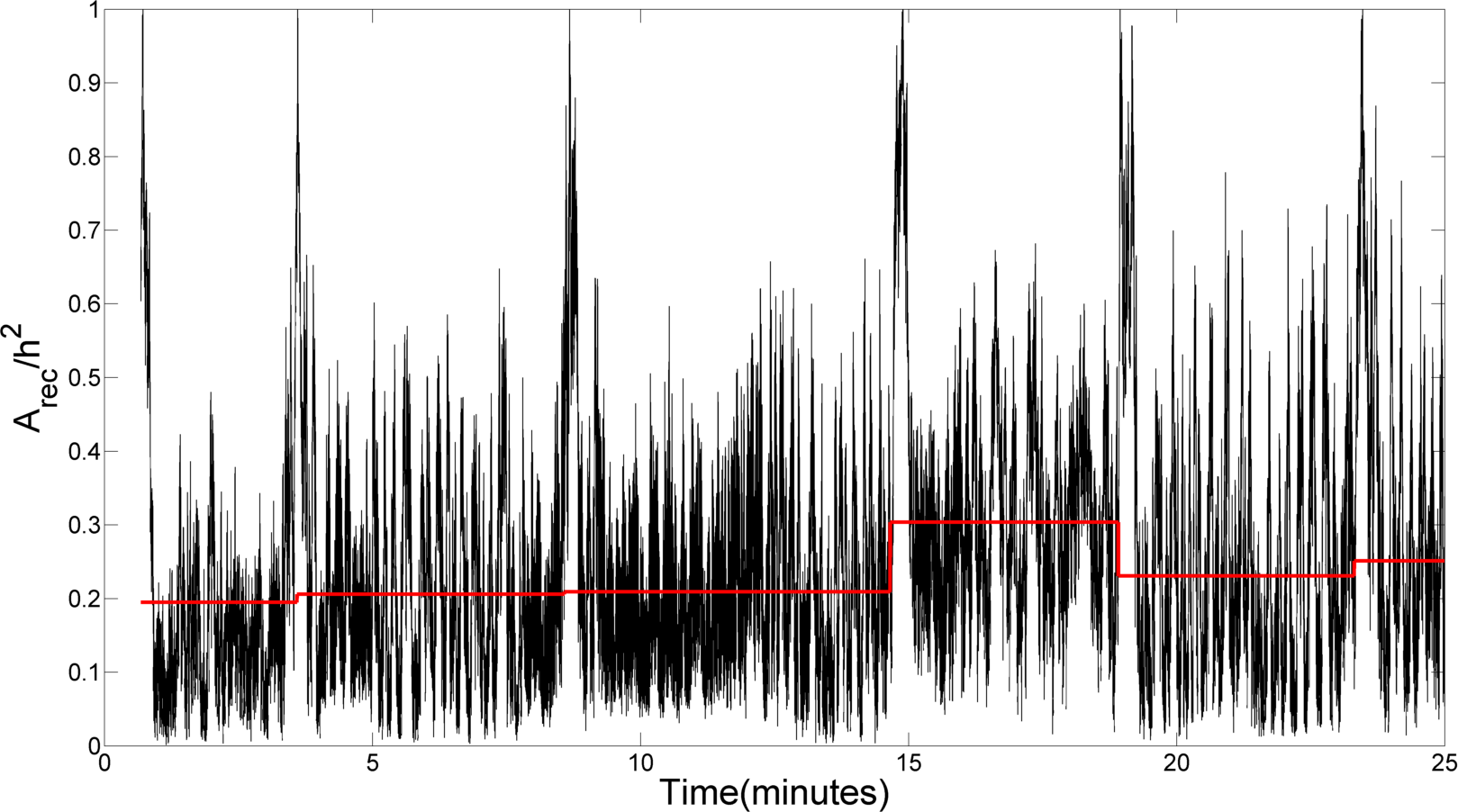}\label{fig:a_recirculation}}
\caption{}
\label{fig:series_freq}
\end{figure}

To demonstrate the efficiency of the frequency-lock approach, the free-stream velocity is randomly varied. Variations in $Re_h$ (based on $u_{check}$) are shown in figure~\ref{fig:u_check}. To ensure strong variations in the KH frequency, a wide range of Reynolds number is explored ($Re_h=1400$ to $2800$). Each time the Reynolds number is changed, the KH frequency is evaluated, as shown on figure~\ref{fig:freq}.

Finally, figure~\ref{fig:a_recirculation} shows the evolution of $A_{rec}(t)$  normalized by $h^2$ as a function of time. The mean value is also computed over each controlled phase (red lines on figure~\ref{fig:a_recirculation}). Re-computation only occurs for major changes in $u_{check}$. What constitutes major changes is up to the user. Because shedding frequency is locked during control to actuation frequency, control is successful even when natural shedding frequency varies slightly.
 Every large peaks in $A_{rec}(t)$ correspond to a re-computation of $f_{KH}$. As a consequence, each peak corresponds to the uncontrolled value associated to the new value of the Reynolds number. When the flow is controlled, reduction in $A_{rec}$ varies between 70 \% to 85 \% when compared to the uncontrolled value.  These values are consistent with those found by \citet{Gautier2013upstream} in open-loop experiments. It clearly demonstrates the robustness of this control strategy based on frequency lock on a natural frequency estimated by real-time instantaneous  optical measurements.

\section{Conclusion}
The experimental study of a frequency lock algorithm was conducted on the separated flow behind a backward facing step. Although basic in nature the control method is able to reliably lower recirculation area. The key is using velocity fields computed from optical data to identify vortex shedding frequency. The shear layer is then periodically excited at the computed frequency. A polling loop continuously checks for changes in the flow state allowing the control to adapt to a randomly changing Reynolds number. Few parameters are required for successful operation and method responsiveness and reliability can be easily and intuitively tweaked based on knowledge of the time scales involved in the relevant flow processes.

\section{Acknowledgments}

The authors wish to thank the DGA and CNRS for their financial support.

\end{document}